# Sliding-induced ferrovalley polarization and possible antiferromagnetic half-metal in bilayer altermagnets

Xin Zhang[1] (张鑫) and Shihao Zhang[2, †] (张世豪)

1   School of Physical Science and Technology, Kunming University, Kunming 650214, China
2   School of Physics and Electronics, Hunan University, Changsha 410082, China
    Corresponding author. E-mail: †zhangshh@hnu.edu.cn

## ABSTRACT

Altermagnets, a newly discovered class of materials, exhibit zero net magnetization while hosting spin-split electronic bands. However, monolayer altermagnets maintain degenerate band gaps at the high-symmetry X and Y points in the Brillouin zone, manifesting a paravalley phase characterized by unpolarized valley states. In this work, we demonstrate that spontaneously broken valley degeneracy can be achieved through interlayer sliding in engineered $M_2A_2B$ and $M_2AA'B$ bilayer altermagnets by first-principles calculations and minimal microscopic model. We propose a promising route to achieve antiferromagnetic half-metal driven by sliding and emergent ferrovalley phase without applied electric field, which is realized in the $V_2SSeO$ engineered bilayer. Our calculations also reveal that $Mo_2O_2O$ exhibits the largest valley splitting gap of ~ 0.31 eV, making it a promising candidate for valley-spin valve devices. Furthermore, band structure calculations on $Mo_2AA'O$ materials demonstrate that increasing the difference in atomic number ($\Delta Z$) between A and A' site atoms effectively enhances valley polarization. This work establishes a novel platform for discovering and controlling ferrovalley states in altermagnetic systems.

## 1 Introduction

Altermagnets (AMs) represent a newly discovered third fundamental magnetic phase beyond ferromagnets and antiferromagnets. It maintains vanishing net magnetization (characteristic of antiferromagnets) while exhibiting momentum-dependent spin-splitting electronic bands (a hallmark of ferromagnets) [1-12]. Consequently, the emergence of altermagnetism has thus sparked extensive investigations, with hallmark phenomena receiving robust confirmation through theoretical calculations and experimental probes: symmetry-breaking lifted Kramers degeneracy [13, 14], Berry-curvature-driven anomalous Hall/Nernst/thermal Hall effects [15-24], crystal-symmetry-enabled nonrelativistic spin currents [25], magneto-optical responses [26, 27], and topologically protected chiral magnon excitations [28, 29]. Simultaneously, altermagnets enable access to enriched physical phenomena, such as valley polarization and multipiezo effects, through diverse external control methods including strain engineering [30-32], electric-field gating [31, 33-36], magnetic-field tuning [18, 30, 37-39], and optical excitation [40]. However, such externally controlled phenomena are inherently volatile, as their effects cease abruptly upon the removal of the applied fields. This necessitates the pursuit of intrinsically nonvolatile material platforms. Inspired by sliding-engineered ferrovalley polarization in $Fe_2MX_4$ systems [41], we constructed interlayer-sliding bilayers from monolayer altermagnetic semiconductors $M_2A_2B$ and $M_2AA'B$ (M = Ti, V, Cr, Mn, Fe, Mo; A, A', B = O, S, Se, Te) [42]. Through first-principles calculations, we identified compounds exhibiting spontaneous ferrovalley polarization.

In the momentum space of many two-dimensional materials, such as graphene and transition metal dichalcogenides (TMDs), the energy extrema often appear in pairs located at specific high-symmetry points of the Brillouin zone, most commonly denoted as the $K$ and $K'$ valleys [43-45]. The electron populations in the two valleys remain equal due to time-reversal symmetry, resulting in a nonpolarized paravalley state. Valley polarization occurs when an external or internal mechanism breaks this symmetry, causing carriers to preferentially occupy one valley, thereby lifting the valley degeneracy [43, 46-48]. This polarized valley degree of freedom can serve as a novel information carrier, analogous to electronic charge or spin. For example, defining the $K$ valley as logic "1" and $K'$ valley as "0" offers a route to encode information, enabling the development of valleytronic devices for data processing, storage, and transmission. Ferrovalley materials represent a distinct class of systems that intrinsically break valley degeneracy by stabilizing one valley at a lower energy than the other, resulting in spontaneous valley polarization without external magnetic field [41, 49]. By combining ferromagnetic order and the valley degree of freedom, such materials provide a promising platform for designing next generation electronic devices with low energy consumption.

In this work, we designed $x$-sliding bilayer altermagnets based on $M_2A_2B$ and $M_2AA'B$ prototypes by first-principles calculations. We propose a minimal microscopic model to describe the ferrovalley states in the bilayer altermagnets. In the family of sliding bilayer altermagnets, $V_2SSeO$ bilayer may achieve antiferromagnetic half-metal induced by sliding operation and emergent ferrovalley mechanism without applied electric field. Furthermore, we performed high-throughout computational screening, and found a series of compounds exhibiting spontaneous valley polarization. And our calculations show that increasing the difference in atomic number ($\Delta Z$) between A and A' site atoms effectively enhances valley polarization. More detailed discussions are presented in the following.

## 2 Computational methods

Our computational framework leveraged the DS-PAW module within the Device Studio platform, implementing density functional theory (DFT) with projector-augmented wave method for sliding configurations [50, 51]. The electron exchange and correlations were described by Perdew-Burke-Ernzerhof (PBE) function. In all calculations, the convergence criteria were set to $10^{-6}$ eV for total energy and 0.05 eV/Å for atomic forces, with a plane-wave energy cutoff of 600 eV. Moreover, a vacuum region of about 20 Å was included to eliminate spurious interactions between periodic layers. The van der Waals interaction between the two layers was accounted for using the DFT-D3 method with Becke-Johnson damping [52]. The Brillouin zone was sampled using a Γ-centered 7 × 7 × 1 Monkhorst-Pack k-point grid [53]. DFT+U calculations were performed with an effective Hubbard parameter $U_{eff}$ = U - J = 3 eV on the d-orbitals of M atoms [54, 55] to include on-site Hubbard interaction.

## 3 Results and discussion

M$_2$A$_2$B and M$_2$AA'B have identical crystal structures, with the A' atom in M$_2$AA'B being chemically equivalent to the A atom. The monolayer M$_2$AA'B systems exhibit the structure shown in Fig. 1(a), the M-B atomic plane is sandwiched between A and A' atomic planes, similar to the monolayer V$_2$STeO [30] and V$_2$SeTeO [56, 57]. The spins of M atoms are denoted by orange arrows, exhibiting A-type antiferromagnetic ordering. Remarkably, specific elemental combinations enable single-layer materials to function as altermagnets [42]. Fig. 1(b) displays AA-stacked bilayer configuration without sliding, which exhibits paravalley characteristic via preserved $\sigma_D$ crystalline mirror symmetry [41]. In order to break $\sigma_D$ symmetry and achieve ferrovalley states, we engineered bilayer systems with orthogonal sliding configurations: x-sliding ($\delta x = a/2$) and y-sliding ($\delta y = b/2$), where one layer undergoes half-lattice-constant displacement relative to the other along corresponding crystal axes, as illustrated in Fig. 1(c) and 1(d).

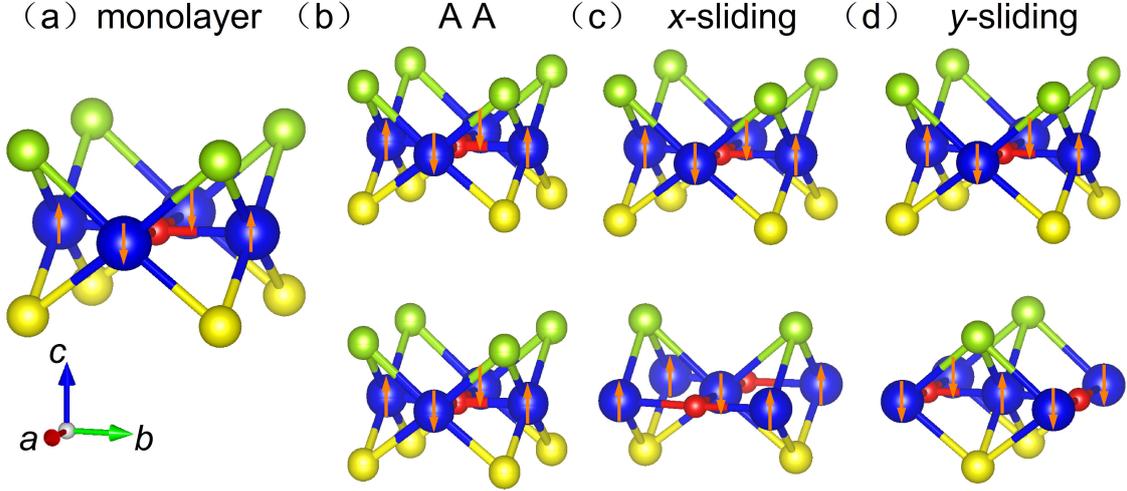

**Fig. 1** Schematic structures of M$_2$A$_2$B and M$_2$AA'B. Blue spheres represent M atoms, green and yellow spheres represent A and A' atoms, and red spheres represent B atoms. (a) Monolayer structure. Antiferromagnetic spins on M atoms are indicated by orange arrows on blue spheres. (b) AA-stacked bilayer structure without interlayer sliding. (c) Bilayer stacking structure with half-lattice-constant sliding along the *x*-axis direction. (d) Bilayer stacking structure with half-lattice-constant sliding along the *y*-axis direction.

We begin discussions about ferrovalley physics with V$_2$SSeO system. Fig. 2 shows the calculated electronic band structure of V$_2$SSeO. As shown in Fig. 2(a), monolayer structure holds direct bandgaps of 214 meV at the X and Y points. Altermagnetism is demonstrated by spin splitting $\sim (k_x^2 - k_y^2)$ in the energy bands, while identical band gaps without spin polarization indicate of paravalley state. Because of $C_{4z}\mathcal{T}$ symmetry, the conduction band minimums at the X and Y points are mainly contributed by $d_{xz}$ and $d_{yz}$ orbitals of different sublattices, respectively. The valence band maximums originate from hybridization of $d_{xy}$ and $p$ orbitals. In the AA-stacking bilayer, the bandgaps are drastically reduced by interlayer interaction as shown in Fig. 2(b).

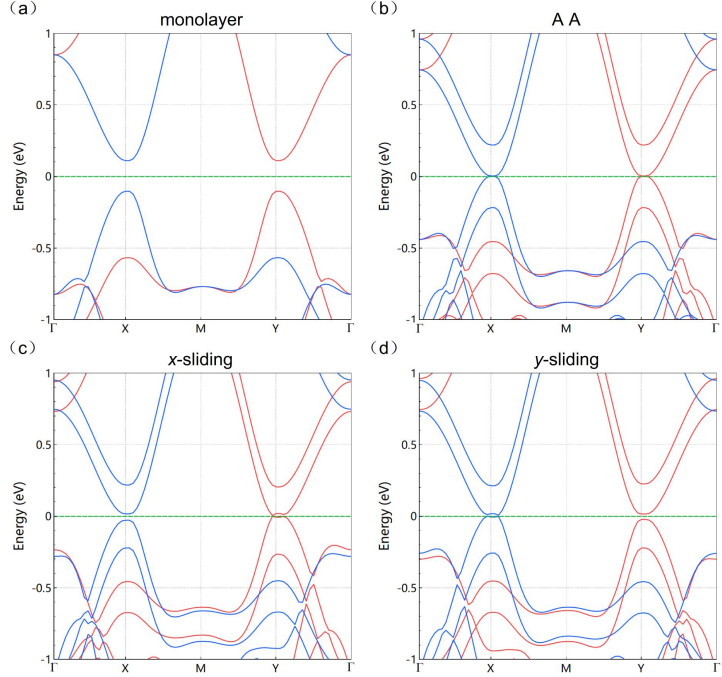

**Fig. 2** Electronic band structures of V$_2$SSeO: (a) monolayer, (b) AA bilayer, (c) *x*-sliding bilayer and (d) *y*-sliding bilayer. The spin-up and spin-down channels are represented by red and blue curves, respectively.

Spin polarization arises in the *x*-sliding bilayer, with the spin-down band at X valley opening a larger gap than the spin-up band at Y valley, yielding a $\Delta_{gap}$ ($\Delta_{gap} = \Delta E_X - \Delta E_Y$) ≈ 18.5 meV, see in Fig. 2(c). The ferrovalley state is antiferromagnetic half-metal, conducting only in one spin channel. Crucially, the observed band inversion in the Y valley provides evidence of nontrivial topological state. Meanwhile, the *y*-sliding bilayer exhibits reversed valley polarization as shown in Fig. 2(d). The two sliding configurations yield antiferromagnetic half-metal with valley polarization with $\Delta_{gap}$ values of opposite signs.

Now we use one minimal model to describe the electronic structures of aforementioned different configurations. We begin with monolayer's effective Hamiltonian [58],

$$H_0 = [\mu + A(\cos k_x + \cos k_y)]\tau_0\sigma_0 + (\cos k_x - \cos k_y)(B\tau_z\sigma_0 + C\tau_0\sigma_z)$$
$$+ t\cos\frac{k_x}{2}\cos\frac{k_y}{2}\tau_x\sigma_0 + [u + D(\cos k_x + \cos k_y)]\tau_z\sigma_z.$$

Here $\tau$ and $\sigma$ denote the Pauli matrices in the sublattice and spin space, respectively. It holds the mirror crystalline symmetry and $C_{4z}\mathcal{T}$ symmetry. In this work, we ignore the spin-orbit coupling for simplicity. $(\cos k_x - \cos k_y)$ term refers to the spin-degenerate bands along $\Gamma - M$ direction. This microscopic model gives out the altermagnetic energy bands as shown in Fig. 3(a). In the AA-stacking configuration, the effective Hamiltonian can be written as $H = H_0 l_0 + t'\tau_0\sigma_0 l_x$, where $l$ is the Pauli matrix defined in the layer space and $t'$ is the interlayer hopping. As shown in the Fig. 3(b), the bilayer's electronic structure evolves into layer-splitting energy bands with smaller bandgap. But in the *x*-sliding or *y*-sliding bilayer structure, mirror symmetry is broken and the spin-degenerate bands along $\Gamma - M$ direction are also split. In the *x*-sliding bilayer configuration, the effective Hamiltonian becomes $H = H_0 l_0 + t'\tau_0\sigma_0 l_x + t''\tau_z\sigma_0 l_0 + t_p\cos k_x \tau_z\sigma_z l_0$. In the *y*-sliding bilayer configuration, the effective Hamiltonian is rewritten as $H = H_0 l_0 + t'\tau_0\sigma_0 l_x - t''\tau_z\sigma_0 l_0 +$

$t_p \cos k_y \tau_z \sigma_z l_0$. As shown in the Fig. 3(c, d), the bilayer's energy bands are driven into ferrovalley states by sliding operation.

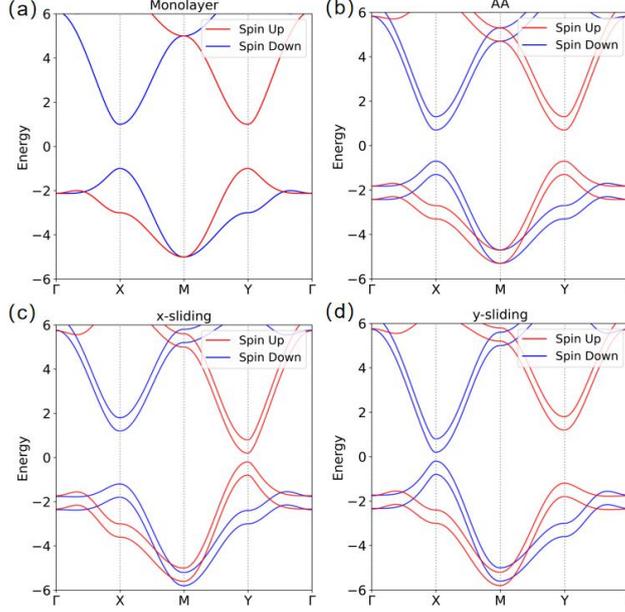

**Fig. 3** Band structures of different configurations. We use the set of parameters: $A = C = -B/2 = D/2 = \mu/2 = t/8 = -u/6 = 0.5$, and $t' = -0.3, t'' = -0.1, t_p = 0.4$.

Given the identical properties of *x*-sliding and *y*-sliding bilayers, we exclusively computed the band structure for the constructed *x*-sliding bilayer in our material simulations. We performed high-throughput calculations to discover more potential altermagnets with remarkable ferrovalley feature. Inspired by recent theoretical work about altermagnet monolayers [42], we focus on the $M_2A_2B$ and Janus $M_2AA'B$ altermagnets because of their possible direct-gap electronic structures. The calculated electronic valley polarizations are summarized in Table 1. As shown in Table 1, many bilayers exhibit metallic electronic structures, because the interlayer coupling reduces the bandgaps. Only few bilayers still keep semiconducting behavior. These semiconducting variants display polarized ferrovalley characteristics, exhibiting $\Delta_{gap}$ values ranging from several to tens of meV. $Mo_2O_2O$ constitutes the sole exception to this trend, possessing a substantially larger $\Delta_{gap}$ of 312.3 meV (Fig. 4), which notably exceeds the reported value of 290 meV for bilayer $Fe_2WTe_4$ [41], thereby establishing an ideal platform for investigating valley spin valves. The screened AM materials consistently exhibit lower energy than their ferromagnetic (FM) counterparts (Table S1), demonstrating the stability of the AM state.

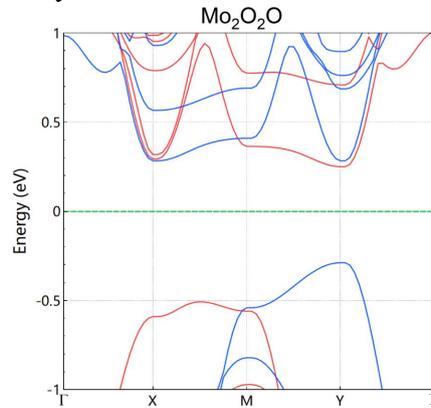

**Fig. 4** The electronic band structures of bilayer $Mo_2O_2O$. The spin-up and spin-down bands are represented by red and blue lines, respectively.

**Table 1** The $\Delta_{gap}$ values (eV) calculated for the constructed bilayer altermagnet materials. Here "/" refers to metallic electronic structure of bilayer altermagnet.

| Materials | $\Delta_{gap}$ (eV) | Materials | $\Delta_{gap}$ (eV) | Materials | $\Delta_{gap}$ (eV) |
|---|---|---|---|---|---|
| $Ti_2O_2O$ | / | $V_2OSSe$ | / | $V_2SSeO$ | 0.0458 |
| $V_2O_2O$ | / | $Ti_2OSeO$ | / | $Cr_2SSeO$ | / |
| $Cr_2O_2O$ | / | $V_2OSeO$ | 0.0018 | $Mo_2SSeO$ | 0.0093 |
| $Mo_2O_2O$ | 0.3123 | $Cr_2OSeO$ | / | $Ti_2SSeSe$ | / |
| $V_2S_2O$ | 0.0170 | $Mo_2OSeO$ | / | $V_2SSeSe$ | / |
| $Cr_2S_2O$ | / | $Fe_2OSeS$ | / | $V_2SSeTe$ | / |
| $Mo_2S_2O$ | / | $Cr_2OSeSe$ | / | $Cr_2STeO$ | / |
| $V_2Se_2O$ | 0.0454 | $Fe_2OSeSe$ | / | $Mo_2STeO$ | 0.0129 |
| $Cr_2Se_2O$ | / | $Mn_2OSeTe$ | / | $Ti_2STeS$ | / |
| $Mo_2Se_2O$ | 0.0041 | $Ti_2OTeO$ | / | $V_2STeS$ | / |
| $Cr_2Te_2O$ | / | $V_2OTeO$ | / | $Ti_2STeSe$ | / |
| $Mo_2Te_2O$ | / | $Cr_2OTeO$ | / | $V_2STeSe$ | / |
| $Ti_2OSO$ | / | $Mo_2OTeO$ | / | $V_2STeTe$ | / |
| $V_2OSO$ | 0.0017 | $Fe_2OTeS$ | / | $Cr_2SeTeO$ | / |
| $Cr_2OSO$ | / | $Fe_2OTeSe$ | / | $Mo_2SeTeO$ | 0.0064 |
| $Mo_2OSO$ | 0.0022 | $Ti_2OTeTe$ | / | $Ti_2SeTeSe$ | / |
| $Cr_2OSS$ | / | $Mn_2OTeTe$ | / | $V_2SeTeTe$ | / |

Fig. 5 presents the electronic band structures of $Mo_2Se_2O$, $Mo_2SeTeO$, and $Mo_2STeO$. All compounds exhibit spin-split bands characteristic of altermagnetism. Crucially, the contrasting bandgap magnitudes between spin-up (spin-down) at X point and spin-down (spin-up) at Y point demonstrate polarized ferrovalley properties. As seen in Table 1, the bandgap follows a clear ascending order from $Mo_2Se_2O$ to $Mo_2SeTeO$ to $Mo_2STeO$, tracking the increasing atomic number difference ($\Delta Z$) between chalcogen sites A and A', which provides a new avenue for discovering materials with enhanced valley polarization.

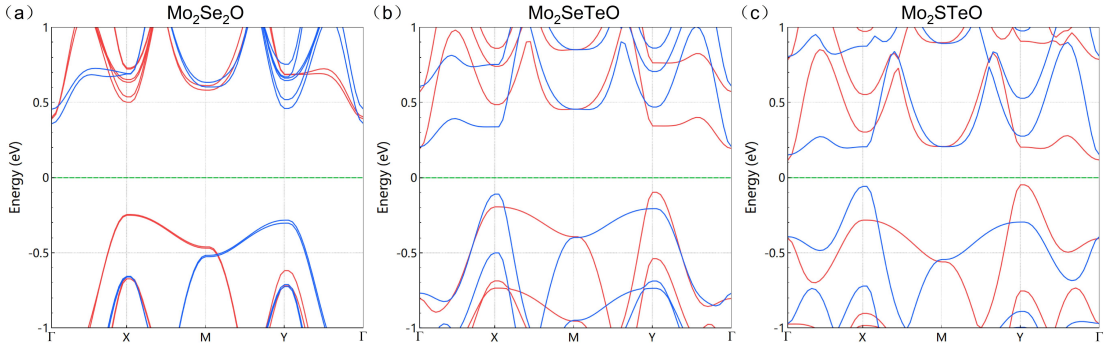

**Fig. 5** The electronic band structures of bilayer (a) $Mo_2Se_2O$, (b) $Mo_2SeTeO$ and (c) $Mo_2STeO$. Spin-up and spin-down channels are represented by red and blue curves, respectively.

We noted that the choice of vdw functionals and Hubbard parameters always influence the calculated bandgaps. However, in this work, we propose a promising route to achieve antiferromagnetic half-metal driven by sliding and emergent ferrovalley phase without applied electric field. By tuning the direction of sliding operation, we can achieve the spin valve with conducting spin-up or spin-down electrons. More importantly, the spin valve has zero net spin moment and then no impact on nearby spintronic components, which promises the robust application in the integrated spintronic devices.

# 4 Conclusions

In conclusion, we designed $x$-sliding bilayer altermagnets based on $M_2A_2B$ and $M_2AA'B$ prototypes. We propose a minimal microscopic model to describe the ferrovalley states in the bilayer altermagnets. In the family of sliding bilayer altermagnets, our calculations reveal that $V_2SSeO$ bilayer may achieve antiferromagnetic half-metal induced by sliding operation and emergent ferrovalley mechanism. Computational screening identified a series of compounds exhibiting spontaneous valley polarization, with $Mo_2O_2O$ showing the largest valley polarization of ~ 0.31 eV. This system offers a promising platform for spin valve applications and provides valuable guidance for further exploration of altermagnetic materials.

**Declarations** The authors declare that they have no competing interests and there are no conflicts.

**Acknowledgements** This work was supported by the National Natural Science Foundation of China (Grant No. 12304217), the Special Basic Cooperative Research Programs of Yunnan Provincial Undergraduate Universities' Association (Grant No. 202401BA070001-013), the Scientific Research Fund of Yunnan Provincial Department of Education (Grant No. 2024J0778), the National Key Research and Development Program of China (Grant No. 2024YFA1410300), the Natural Science Foundation of Hunan Province (Grant No. 2025JJ60002), and the Fundamental Research Funds for the Central Universities from China (Grant No. 531119200247). We gratefully acknowledge HZWTECH for providing computation facilities.